# Anomalous Ferromagnetism in TbMnO$_3$ Thin Films

B. J. Kirby,[1,a)] D. Kan,[2] A. Luykx,[2] M. Murakami,[2] D. Kundaliya,[2] and I. Takeuchi[2]

[1]*Center for Neutron Research, National Institute of Standards and Technology, Gaithersburg, Maryland, 20899, USA*

[2] *Department of Materials Science and Engineering, University of Maryland, College Park, Maryland, 20742, USA*

ABSTRACT

Magnetometry, x-ray, and neutron scattering have been used to study the structural and magnetic properties of a TbMnO$_3$ thin film grown on a [001] SrTiO$_3$ substrate by pulsed laser deposition. Although bulk TbMnO$_3$ is a low temperature antiferromagnet, magnetometry measurements indicate the presence of low temperature ferromagnetism. Depth profiling by x-ray and polarized neutron reflectometry reveals a net sample magnetization that is commensurate with the film thickness, indicating that the observed ferromagnetism is not due to an altered surface phase (such as Mn$_3$O$_4$), or external impurities that might give rise to an artificial magnetic signal. Instead, these results show that the ferromagnetism is an intrinsic property of the TbMnO$_3$ film.

Multiferroic materials, which feature simultaneous coupled ordering of magnetic moments and electric polarization, are the subject of widespread research interest. Multiferroics hold the promise of novel devices in which the the magnetic properties are electrically controlled, and vice versa. One example is that of an antiferromagnetic multiferroic thin film used to produce a preferred magnetization orientation in an adjacent ferromagnetic thin film via exchange bias [1,2] coupling. For such a structure, it is possible to alter the nature of the exchange coupling (and therefore the magnetic orientation of the ferromagnet) by applying an electric field to change the interfacial magnetic domain structure of the multiferroic antiferromagnet. For example, it has recently been demonstrated that the magnitude of the low temperature exchange bias field in $YMnO_3$ / permalloy multilayers changes as a function of the electric field applied during cooling.[3]

To further pursue the technological possibilities of exchange biasing multiferroics, other materials need to be explored. In bulk, $TbMnO_3$ is a multiferroic material which in bulk exhibits a complex magneto-electric phase diagram, including antiferromagnetism below 30 K - making it a candidate for exchange-bias applications.[4-6] However, scaling down manganite crystals into thin film form causes strain that can drastically affect the magnetic properties,[7] and relatively little work has been done to study the magnetic properties of $TbMnO_3$ thin films. As a first step towards determining the utility of $TbMnO_3$ for exchange-bias applications, we have produced a thin film of the orthorhombic phase of $TbMnO_3$, and studied it with magnetometry, x-ray, and neutron scattering measurements. These measurements show that the $TbMnO_3$ film exhibits an intrinsic *ferro*magnetic order.

A $TbMnO_3$ film was laser ablated from a stoichiometric $TbMnO_3$ target, onto a [001] oriented $SrTiO_3$ substrate. Figure 1 shows results of $\theta$-$2\theta$ x-ray diffraction (XRD) scans, performed with a commercial diffractometer.[8] The fabricated film exhibits the [00$l$] peak of the perovskite structure, with an out-of-plane lattice parameter $c = 0.372$ nm. No peaks identifiable as originating from additional phases were observed. X-ray reciprocal space mapping measurements around the SrTiO$_3$ [103] Bragg reflection (not shown) reveal that the in-plane lattice parameter of the TbMnO$_3$ film closely matches that of the cubic SrTiO$_3$ substrate (0.391 nm). Bulk TbMnO$_3$ has an orthorhombic perovskite structure which (in the pseudo-cubic convention) has lattice parameters of 0.393 nm in-plane, and 0.370 nm out-of-plane.[9] Therefore, we conclude the fabricated film is compressively strained by the substrate, resulting in the tetragonally distorted orthorhombic phase of TbMnO$_3$.

Net magnetization measurements of the sample were performed using superconducting quantum interference device (SQUID) magnetometry. Figure 2 shows the field $H$ dependent magnetic moment $m$ at temperature $T = 5$ K after cooling from 100 K in zero field.[10] Hysteretic behavior is observed, clearly indicating ferromagnetism - a drastic departure from the antiferromagnetic order observed for bulk TbMnO$_3$. $m(T)$ in a 1.5 T field, is shown in the Fig. 1 inset, and reveals that a net magnetization persists above 100 K. However, standard magnetometry measurements alone cannot determine whether the observed ferromagnetism is present throughout the film, or if it is instead due to a departure from the TbMnO$_3$ stoichiometry in a small region of the film.

To determine if a net magnetization is truly intrinsic to the TbMnO$_3$ film, we determined the structural and magnetic depth profiles via x-ray and polarized neutron reflectometry [11] (XRR and PNR, respectively) measurements at the NIST Center for Neutron Research. XRR measurements were conducted at room temperature with wavelength $\lambda = 0.154$ nm x-rays (Cu$_{k\alpha}$), and PNR measurements were conducted at cryogenic temperatures with $\lambda = 0.475$ nm cold neutrons. The neutron beam was polarized by Fe/Si supermirror and a Mezei spin-flipper to be alternately spin-up (+) or spin-down (-) relative to an in-sample-plane magnetic field $H$, and was incident on the sample. The reflected beam was intercepted by an analyzer supermirror/flipper assembly so that all four PNR cross sections (non spin-flip: $R^{++}$, $R^{--}$, and spin-flip: $R^{-+}$, $R^{+-}$) could be measured with a $^3$He pencil detector.

In reflectometry, depth ($z$) dependent sample properties are determined from the scattering vector ($Q$) dependent reflectivities.[12] The x-ray and neutron beams are both highly penetrating, and probe the entirety of the TbMnO$_3$ thin film. XRR is sensitive to the sample's x-ray scattering length density $\varrho_X(z)$ (a function of electron density),[13] while PNR is sensitive to the sample's nuclear scattering length density $\varrho_N(z)$ and volume magnetization $M(z)$.[14,15] Specifically, the non spin-flip scattering $R^{++}$ and $R^{--}$ depend on $\varrho_N(z)$ and the component of $M(z)$ parallel to $H$, while the spin-flip scattering $R^{+-}$ and $R^{-+}$ depend solely on the component of $M(z)$ perpendicular to $H$. Thus, the structural and magnetic depth profiles of thin film structures can be determined by model-fitting XRR and PNR spectra. For this work, model fitting was done using the NIST GA_REFL software package.[16]

Fitted Cu$_{k\alpha}$ XRR data are shown in Figure 3. Clear oscillations are observed, indicating a strong sensitivity to the interface between the TbMnO$_3$ film and the SrTiO$_3$ substrate. PNR measurements of the same sample were conducted in a 0.55 T field, after cooling the sample from above 290 K to 6 K in a 5.5 T field. The spin-flip channels were measured only at 6 K, and scattering was found to be at background levels, indicating no detectable magnetic component perpendicular to $H$. The fitted non spin-flip scattering data are shown in Figure 4. At 6 K (a), $R^{++}$ and $R^{--}$ exhibit clear oscillations that are roughly 180 degrees out of phase from one another. In the absence of nuclear spin polarization,[18] the only way these two spin states can differ is if the sample possesses a net magnetization parallel to $H$. Further, the common frequency of the $R^{++}$ and $R^{--}$ oscillations is essentially the same as that of the XRR oscillations shown in Figure 3, suggesting that the magnetized film thickness is similar to the total film thickness. At 100 K (b), the splitting between $R^{++}$ and $R^{--}$ persists, indicating that the sample is still magnetized. The higher frequency oscillations observed in panel a) and Fig. 3 are much weaker at 100 K. This indicates that while neutrons are very sensitive to the *magnetic* interface between the film and the substrate, they are not very sensitive to the *nuclear* interface. With the high frequency oscillations damped, a very weak lower frequency oscillation becomes apparent, suggesting some depth dependence of $\rho_N$ within the TbMnO$_3$ film.

Results of model fitting the XRR and PNR data are shown in Figure 5. The structural profiles are shown in panel a) ($\rho_N$ and both the real and imaginary components of $\rho_X$). The x-ray profiles feature a distinct substrate/film interface, and show that the TbMnO$_3$ film is 70 nm thick. The nuclear profile features a slight decrease in $\rho_N$ from top to bottom of the film, possibly due to a small variation in film strain. The magnetic profiles are shown in Fig 5 b), and confirm that the TbMnO$_3$ possess a net magnetization parallel to $H$ that is essentially constant across the entirety of the 70 nm film. Given the lattice parameters determined from XRD, the profiles shown in Fig. 5 correspond to magnetic moments of 0.6 and 0.1 $\mu_B$ per unit cell at 6 K and 100 K, respectively.

These results show that the ferromagnetism observed with SQUID is an intrinsic property of the TbMnO$_3$ thin film. While we cannot completely rule out a uniform distribution of impurities as the origin of ferromagnetism (magnetic clusters, for example), we have definitively shown that the ferromagnetism is not due to surface impurities. Specifically, observation of a uniform sample magnetization at 100 K confirms that a surface layer of ferrimagnetic Mn$_3$O$_4$ ($T_C$ = 42 K) [20] is not the source of the ferromagnetic signal observed with SQUID. Instead, it seems that the ferromagnetic behavior is indeed an intrinsic property of strained orthorhombic TbMnO$_3$ films. Very recent unpublished work by Rubi *et al*. also shows that thin TbMnO$_3$ films on SrTiO$_3$ substrates can exhibit ferromagnetism, and they present evidence suggesting that the ferromagnetism arises from in-plane compressive strain. [21] This is a likely explanation for the ferromagnetism we have observed, as our film is also compressively strained in-plane. Determining how to tune the growth parameters to produce fully antiferromagnetic TbMnO$_3$ films for exchange bias applications remains an important avenue of future research. In summary, our results reveal additional complexity to the already rich TbMnO$_3$ magneto-electric phase diagram, and have important implications for the role of TbMnO$_3$ as a biasing multiferroic antiferromagnet.

This work was supported by NSF-MRSEC under grant No. DMR 0520471, NSF DMR 0603644, and ARO W9IINF-07-1-0410. We thank J. A. Borchers of NIST for valuable discussions.

REFERENCES

[1] W. H. Meiklejohn and C. P. Bean, Phys. Rev. **105**, 904 (1957).

[2] J. Nogues and I. K. Schuller, J. Magn. Magn. Mater. **192**, 203 (1999).

[3] V. Laukhin, V. Skumryev, X. Marti, D. Hrabovsky, F. Sanchez, M. V. Garcia-Cuenca, C. Ferrater, M. Varela, U. Luders, J. F. Bobo, and J. Fontcuberta, Phys. Rev. Lett. **97**, 227201 (2006).

[4] T. Kimura, G. Lawes, T. Goto, Y. Tokura, and A. P. Ramirez, Phys. Rev. B **71**, 224425 (2005).

[5] T. Goto, T. Kimura, G. Lawes, A. P. Ramirez, and Y. Tokura, Phys. Rev. Lett. **92**, 257201 (2004).

[6] T. Kimura, T. Goto, H. Shintani, K. Ishizaka, T. Arima, and Y. Tokura, Nature **426**, 55 (2003).

[7] D. Rubi, Sriram Venkatesan, B. J. Kooi, J. Th. M. De Hosson, T. T. M. Palstra, and B. Noheda1, *Physical Review B* **78**, 020408(R) (2008).

[8] $c = \lambda / 2\sin\theta$, where $\lambda$ is the x-ray wavelength of 0.154 nm, and $\theta$ is the scattering angle.

[9] J. A. Alonso, M. J. Martinez-Lope, M. T. Casais, and M. T. Fernandez-Diaz, *Inorganic Chemistry* **39**, 917 (2000).

[10] Magnetometry measurements of similarly prepared TMO samples show that field cooling vs. zero field cooling has no observable effect on the resulting hysteresis loop.

[11] See http://www.ncnr.nist.gov/instruments/ng1refl.

[12] For monochromatic radiation (as for the cases described here), $Q = 4\pi \sin\theta / \lambda$, where $\theta$ is the scattering angle, and $\lambda$ is the radiation wavelength.

[13] L. G. Parratt, *Physical Review* **95**, 359 (1954).

[14] G. P. Felcher, Phys. Rev. B **29**, 1268 (1984).

[15] C. F. Majkrzak, Physica B **221**, 342 (1996).

[16] P. A. Kienzle, M. Doucet, D. J. McGillivray, K. V. O'Donovan, N. F. Berk, and C. F. Majkrzak, see http://www.ncnr.nist.gov/reflpak for documentation.
7

FIGURES

Figure 1: The SrTiO3 and TbMnO3 [001] reflections observed using x-ray diffraction. Error bars indicate +/- 1 sigma.

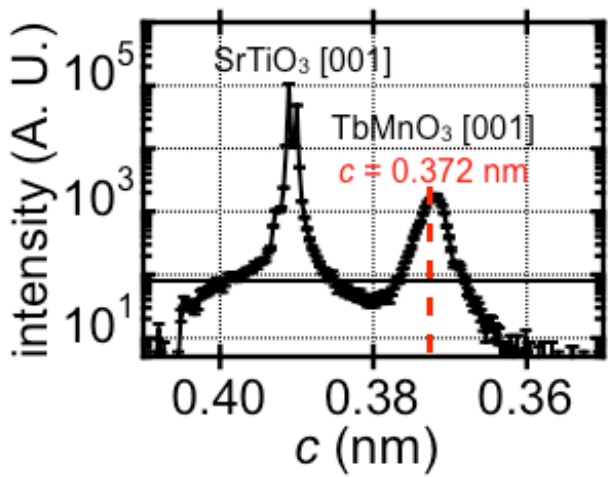

Figure 2: SQUID magnetometry results. Ferromagnetism is observed at 5 K, and a net magnetization persists above 100 K (inset).

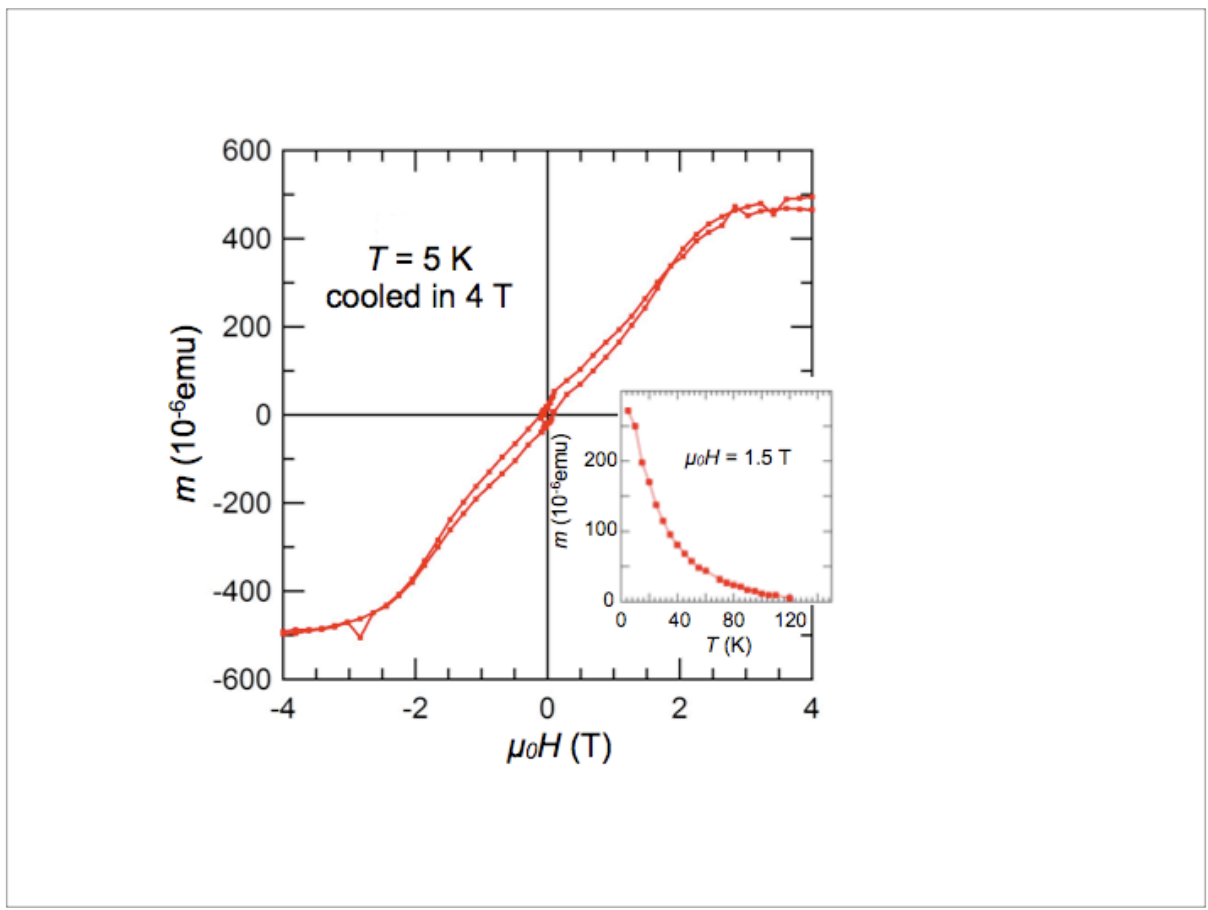

Figure 3: Fitted XRR data for a typical TbMnO3 film. Some higher Q data is omitted for clarity. Error bars indicate +/- 1 sigma.

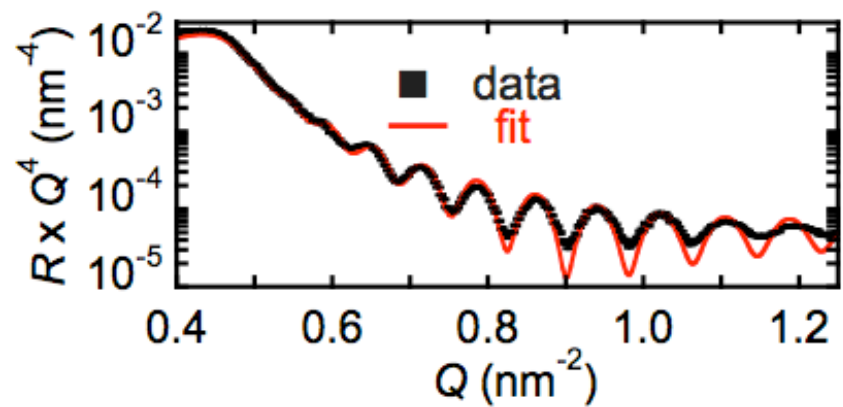

Figure 4: Fitted PNR. Spin-dependent oscillations are present at 6 K (a), and at 100 K (b, inset). Error bars indicate +/- 1 sigma.

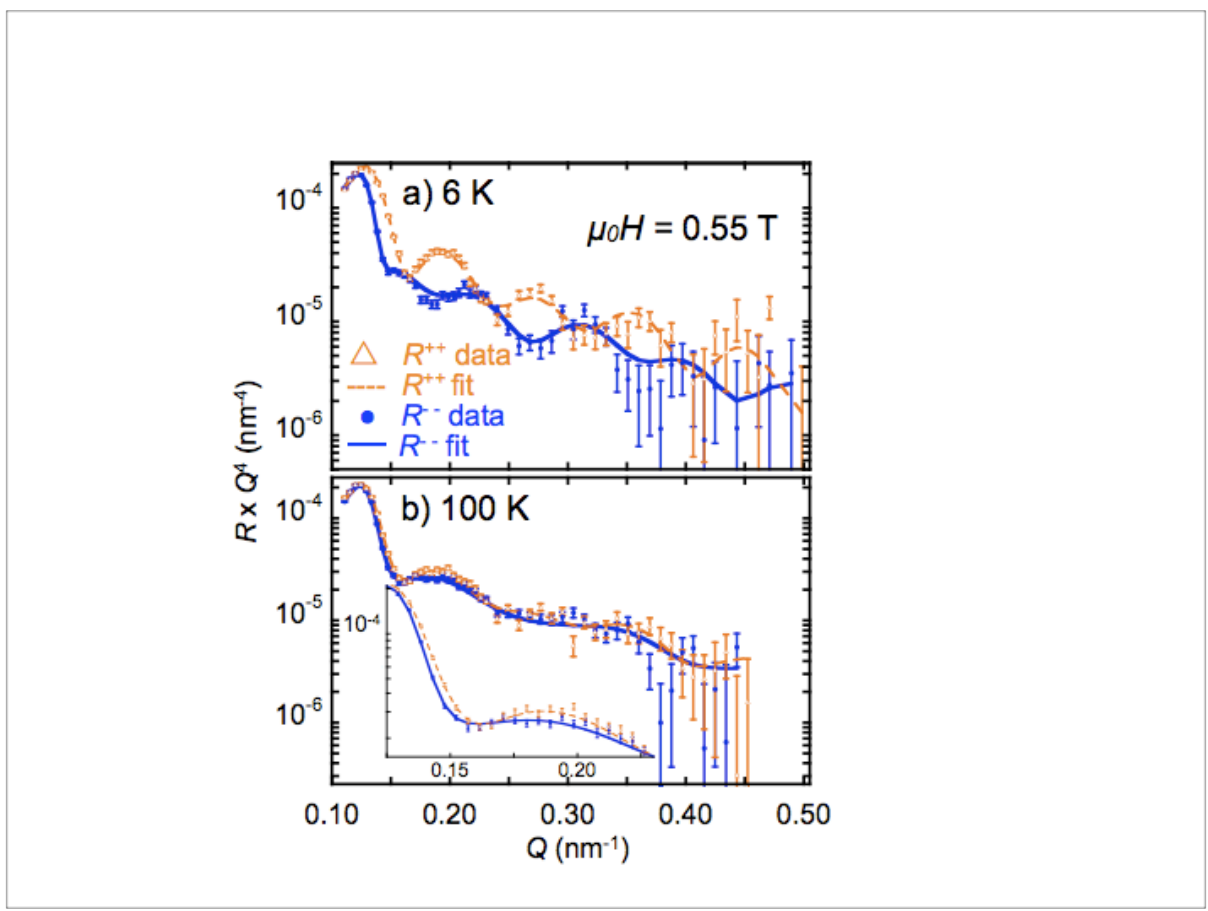

Figure 5: Models used to fit the data in Fig. 2-3. a) Structural profiles: real and imaginary components of the x-ray scattering length density, and the nuclear scattering length density (multiplied 10). b) Magnetic profiles at 6 K and 100 K.

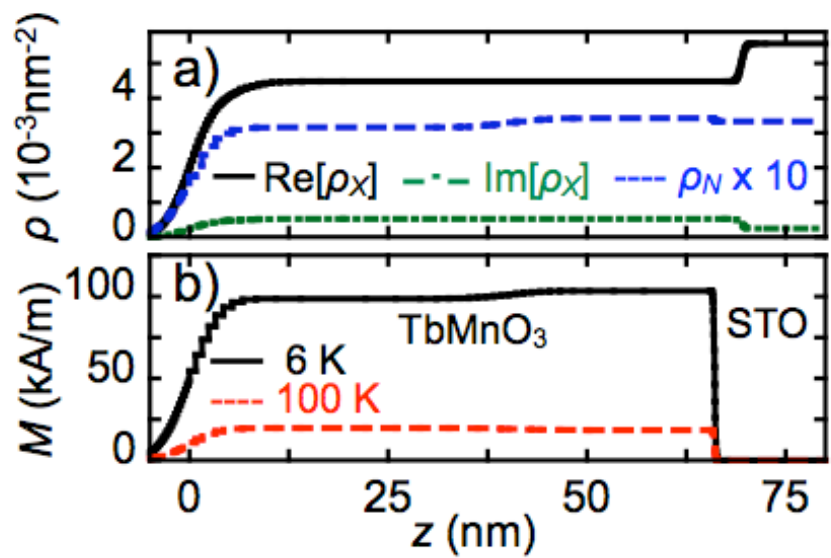